\documentclass[12pt]{article}
\usepackage{amsmath,amssymb,graphicx,mathrsfs,hyperref,slashed,setspace}

\usepackage[title]{appendix}
\usepackage{float}
\hypersetup{
	colorlinks=true,
	linkcolor=blue,
	filecolor=black,
	urlcolor=black,
	citecolor=blue,
}

\newcommand{\hide}[1]{}

\newcommand{\be}{\begin{equation}}
\newcommand{\ee}{\end{equation}}
\newcommand{\bea}{\begin{eqnarray}}
\newcommand{\eea}{\end{eqnarray}}

\def\({\left(} \def\){\right)}

\begin{document}
\title{\vspace{-1.8in}\hspace{-.8in}
{Black holes as frozen stars: Regular interior geometry}}
\author{\large Ram Brustein${}^{(1)}$,  A.J.M. Medved${}^{(2,3)}$, Tom Shindelman${}^{(1)}$,
Tamar Simhon${}^{(1)}$
\\
\vspace{-.5in} \hspace{-1.5in} \vbox{
\begin{flushleft}
 $^{\textrm{\normalsize
(1)\ Department of Physics, Ben-Gurion University,
   Beer-Sheva 84105, Israel}}$
$^{\textrm{\normalsize (2)\ Department of Physics \& Electronics, Rhodes University,
 Grahamstown 6140, South Africa}}$
$^{\textrm{\normalsize (3)\ National Institute for Theoretical Physics (NITheP), Western Cape 7602,
South Africa}}$
\\ \small \hspace{-0.37in}
   ramyb@bgu.ac.il,  j.medved@ru.ac.za, tomshin@post.bgu.ac.il,  simhont@post.bgu.ac.il
\end{flushleft}
}}
\date{}
\maketitle

\setstretch{1.3}
\begin{abstract}

We have proposed a model geometry for the interior of a regular  black hole mimicker, the frozen star, whose most startling feature is that each spherical shell in its interior is a surface of infinite redshift. The geometry is a solution of the Einstein equations which is sourced by an exotic matter with maximally negative radial pressure. The frozen star geometry was previously presented in singular coordinates for which $-g_{tt}$ and $g^{rr}$ vanish in the bulk and connect smoothly to the Schwarzschild exterior. Additionally, the geometry was mildly singular in the center of the star. Here, we present regular coordinates for the entirety of the frozen star. Each zero in the metric is replaced with a small, dimensionless parameter $\varepsilon$; the same parameter in both  $-g_{tt}$ and $g^{rr}$ so as to maintain maximally negative radial pressure. We also regularize the geometry, energy density and pressure in the center of the star in a smooth way. Our initial analysis uses Schwarzschild-like coordinates and applies the Killing equations to show that an infalling, point-like object will move very slowly, effectively sticking to the surface of the star and never coming out. If one nevertheless follows the trajectory of the object into the interior of the star, it moves along an almost-radial trajectory until it comes within a small distance from the star's center. Once there, if the object has any amount of angular momentum at all, it will be reflected outwards by a potential barrier onto a different almost-radial trajectory. Finally, using Kruskal-like coordinates, we consider the causal structure of the regularized frozen star and discuss its  $\varepsilon\to 0$ limit, for which the geometry degenerates and becomes effectively two dimensional.
\end{abstract}
\maketitle

\newpage

\section{Introduction}
\setstretch{1.5}

The term black hole (BH) has been universally adopted to describe the final state of matter after it gravitationally collapses and is meant to reflect the singular nature of its classical solution; something that had long been suspected \cite{ray,komar,Buchdahl,chand1,chand2,bondi} but only proven in the classic works of Penrose and Hawking \cite{PenHawk1,PenHawk2}. On the other hand, given that quantum theory is expected to resolve all singularities, a more appropriate term for the final state might be a frozen star, as was first coined in   \cite{Ruffini:1971bza}. This is because  of the infinite time for gravitational collapse to transpire from the perspective of an external observer and because deviations away from the static Schwarzschild geometry decay exponentially fast on a scale that is set by the light-crossing time. In other words, the collapsing matter configuration can, for all practical purposes, be regarded as frozen in time. In this spirit, we have adopted the name ``frozen star'' for our own model of the final state of matter a long time after it collapsed.

As for quantum mechanics' role as the guardian of regularity, a common expectation is that quantum effects at the Planck scale will be sufficient for this purpose. Although this idea cannot be ruled out in general, there are strong indications to the contrary in the context of BH singularities. First,  a seemingly necessary condition  for evading the singularity theorems \cite{PenHawk1,PenHawk2} and the closely related  ``Buchdahl-like''  bounds \cite{Buchdahl,chand1,chand2,bondi} is that  the geometry is sourced by matter having the most negative radial pressure  that is permitted by causality, $\;p_r=-\rho\;$, all the way out to the surface of the star \cite{bookdill}. This property was an essential ingredient in the black star model \cite{barcelo}, the gravastar model \cite{MMfirst,MM} and  a hybrid of the two \cite{CR}.  Furthermore, if one also considers the emitted Hawking radiation from  a regularized BH mimicker, what is found is an untenable violation of energy conservation  when the scale of resolution is parametrically smaller than that  of the Schwarzschild radius $R_S$. Indeed, in this case, the emitted energy of Hawking particles will greatly exceed the original mass of the collapsing matter \cite{frolov,visser}.
The natural conclusion is that a regularized BH mimicker is required to have deviations from classical general relativity that  extend throughout the object's interior. For a comprehensive list and extensive discussion on compact objects that are meant to mimick BHs, see \cite{carded}.

One such BH mimicker, known as the collapsed polymer model, was proposed by two of the current authors \cite{strungout} on the basis that the object's interior should be filled up with a maximally entropic fluid  \cite{inny}, a state that is best described by a Hagedorn phase of  highly entropic stringy matter \cite{AW,SS,LT,HP,DV}. Utilizing, in particular, a collection of long, closed, interacting strings,  we were able to replicate all known features of Schwarzschild BHs \cite{emerge} and make a number of novel predictions about the non-equilibrium physics \cite{ridethewave} that could possibly be tested via
the observation of gravitational-wave emissions during binary-BH mergers \cite{spinny,collision,QLove,CLove}.

The problem with the polymer model is that its highly quantum interior cannot be  described by a semiclassical metric.
The way out of this conundrum is to identify a classical geometry that maintains many of the same  characteristics as the polymer model \cite{bookdill} or, put  differently, understand how the polymer BH would be viewed by someone who is ignorant about the microscopic nature of its interior or, more so, someone who is determined --- by hook or by crook --- to forgo quantum mechanics in her picture of  gravitational collapse \cite{BHfollies}.

The polymer's geometric proxy, the frozen star, was assumed initially to have the following prominent features in terms of the energy density $\rho$, the radial pressure $p_r$ and the transverse pressure $p_{\perp}$:
\begin{enumerate}
\item
It has maximally negative radial pressure, $\;p_r=-\rho\;$,  which  implies a specific geometry,  $\;f(r)\equiv -g_{tt}=g^{rr}\;$.
\item
It has vanishing transverse pressure, $\;p_{\perp}=0\;$.
\item
The interior metric, which is defined for $\;r\le R\;$ with $\;R\simeq R_S\;$, has the same form as that of the Schwarzschild horizon,  $\;f(r)=0\;$ everywhere except for a thin layer at the outer surface \cite{popstar} and a small region surrounding the center (see below). In spite of this, it is regular throughout the interior
\item
It is ultra-stable against perturbations \cite{bookdill,popstar}.
\end{enumerate}

In previous papers,  the condition $\;f(r)=0\;$ has been strictly enforced in the bulk of the interior. However, this geometry has apparently singular coordinates which makes it hard to deduce its physical consequences. The main objective of our current paper is to study a more accessible geometry by relaxing this condition, setting  $\;f(r<R)= 1-v^2\;$  for $\;v^2\lesssim 1\;$, with $\;\varepsilon=1-v^2\ll 1\;$ as the small parameter in this model.   Such a geometry has been referred
to as a hedgehog compactification  elsewhere in a cosmological context  \cite{guendel2,guendel1}. Notice that we are not relaxing the condition
$\;-g_{tt}=g^{rr}\;$ (including in the outer layer and the central region), as  it this choice that ensures $\;p_r=-\rho\;$. Also note that, as long as $f(r)$
is a constant, the relaxing of (3) has no bearing on (2),  $\;p_{\perp}=0\;$ still holds.

The  remainder of the paper proceeds as follows: First,  we utilize the  Schwarzschild-like
(or hedgehog) coordinate system along with the Killing equations to show that, from an external perspective, an infalling particle would take a very long time, $\;R/\sqrt{\varepsilon}\gg R\;$, to re-emerge from the frozen star. Here, it is also shown that the gravitational potential forms an infinite
angular momentum barrier near the center of the frozen star at  $\; r\sim R \sqrt{\varepsilon}\;$ for any particle with non-zero angular momentum.  The barrier implies that
almost all of the infalling particles avoid the origin.

Next, in Section~3, we discuss the causal structure of the frozen star by introducing Kruskal-like coordinates, which cast the metric in a form that is manifestly regular even in  the $\;\varepsilon \to 0\;$ limit. This form is useful for better understanding this limiting case, as well as for understanding  the corresponding  Penrose Diagrams.

As a prologue to discussing Section~4,
let us first mention that  the energy density and most curvature invariants formally diverge in the combined $\;r\to 0\;$ and $\;\varepsilon\to 0\;$ limits. The divergence is rather mild as the total mass in the central region is finite and small. Nevertheless, as a  goal of this paper is to present a completely regular metric everywhere in space, we need to regularize the geometry in the region close to the center.  The regularization procedure  is summarized in Section~4 of the main text and closely follows our analysis in \cite{popstar} for the outer layer. Technical details about the regularization are presented in an appendix.

The paper ends with a  short comment about stability,  a brief overview and the aforementioned Appendix.

\subsection*{Conventions}

We assume a spherically symmetric and static background spacetime
with  $\;D=3+1\;$ spacetime  dimensions, but similar results will persist for any $\;D>3\;$. All  fundamental constants  besides Newton's constant $G$ are set to unity throughout except when included for clarity and  $\;8\pi G=1\;$ is used  in Section~4 and the Appendix.  A prime indicates a radial  derivative.

\section{The interior geometry}

Let us begin here with a review and then present the regularized version of the interior geometry of the frozen star.
Here, we will ignore the thin layer near the outer surface which  must be  modified to ensure that  the solution can be
matched smoothly to the Schwarzschild exterior \cite{popstar} and also the small region near the center which
must be regularized  to ensure  that all densities remain finite (see Section~4).

A static and spherically symmetric
line element is assumed,
\be
ds^2\; =\; -f(r) dt^2 + \frac{1}{{\widetilde f}(r)} dr^2 + r^2 (d\theta^2+ sin^2 \theta d\phi^2)\;.
\label{linement}
\ee
It is further assumed that the radial pressure is maximally negative,  $\;p_r=-\rho\;$,
the transverse components  $p_{\perp}$ are initially unspecified and all of the off-diagonal
components are vanishing.
Under these conditions,  Einstein's equations reduce to
\begin{align}
\label{E1}
\left(r{\widetilde f}\right)'  & \;=\; 1-8\pi G\rho r^2\;, \\
\label{E2}
\left(rf\right)''\; & \;=\; 16\pi G r p_{\perp}\;.
\end{align}
where  $\;f={\widetilde f}\;$ due to the maximally negative pressure.
This can all be combined into a single equation of the form
\be
\label{E3}
\left(\rho r^2\right)'  \;=\; - 2 r p_{\perp}\;,
\ee
which also follows from the stress-tensor conservation equation.

Next, defining the mass function,
\be
\label{mofr}
m(r)\;=\;4\pi G\int\limits_0^r dx\, x^2 \rho(x)\;\;\; {\rm for} \;\;\; r\leq R\;,
\ee
we find that
\be
f(r)\;=\; {\widetilde f}(r) \;=\; 1- \frac{2G m(r)}{r}\;.
\ee
The functional form of $m(r)$ determines the geometry. For example, if
$m$ is chosen so that $\rho$ is constant, the result is the gravastar model.

The frozen star corresponds to the choice
\be
\;m(r)=r/2G\;
\label{mR}
\ee
throughout the interior.
This in turn implies that $\;f=0\;$ and the matter densities take the forms
\bea
\label{polyrho}
8\pi G  \rho &=& \frac{1-(rf)'}{r^2} \;=\;  \frac{1}{r^2}\;, \\
\label{polypr}
8\pi G  p_r &=& -\frac{1-(rf)'}{r^2} \;=\; -\frac{1}{r^2}\;, \\
\label{polypt}
8\pi G  p_\perp &=& \frac{(rf)''}{2r}\;=\; 0\;.
\eea
The radius of the star $R$ in this case is exactly the Schwarzshild radius $\;R_S=2GM=2G m(R)$ and the exterior geometry is exactly the Schwarzschild geometry.~\footnote{Recall that we are ignoring the thin region near the surface of the star where its density decreases smoothly to zero to match the Schwarzschild exterior.}

In this paper, we follow a somewhat different route and rather set $\;2Gm(r)= r\upsilon^2\;$, so
that $\;f=1-\upsilon^2=\varepsilon\ll 1\;$. Then, in addition to having a regular geometry almost everywhere in the
bulk~\footnote{The exception is a small region near the center, which will be discussed separately.}, the coordinates are also regular.

The line element inside the star is then the following:
\be
ds^2 \;=\; -\varepsilon dt^2 + \frac{1}{\varepsilon} dr^2 + r^2 d\Omega_2^2\;.
\label{HHog}
\ee
As already mentioned, this metric was first introduced in \cite{guendel2,guendel1} in a cosmological context. The quantity $\upsilon$ has the dimensionality of a velocity, and indeed, we will find that this is the proper radial velocity inside the frozen star of massive objects which start at rest at infinity (see Eq.~(\ref{v4mass})).

The previous densities in Eqs.~(\ref{polyrho}), (\ref{polypr}) are only slightly modified,
\bea
\label{polyrho2}
8\pi G  \rho &=&  \frac{\upsilon^2}{r^2}\;, \\
\label{polypr2}
8\pi G  p_r &=&  -\frac{\upsilon^2}{r^2}\;, \\
\label{polypt2}
8\pi G  p_\perp &=&  0\;,
\eea
and are recovered in the limit $\;\upsilon^2\to 1\;$ or, equivalently, $\;\varepsilon\to 0\;$.

The radial position of the surface of the star $R$  for a star of mass $M$  shifts out by a parametrically small amount from its Schwarzschild value,
\begin{equation}
R\;=\;\frac{2GM}{\upsilon^{2}}\;=\;\frac{2GM}{1-\varepsilon}\;\approx\;2GM\left(1+\varepsilon\right)\;.
\label{eq:modradius}
\end{equation}
Again, for $\;r > R\;$, the geometry is exactly that of  Schwarzschild  for  a star of mass $M$.

Here, we do not discuss explicitly the transitional layer near the surface of the star, in which the density continuously decreases to zero to match  the outer Schwarzschild geometry. We have verified that the relevant analysis in \cite{popstar} can be extended in a straightforward manner to the current model. In both cases, the position of the outer surface depends on the width of the translational layer and, for the current case, the outermost surface is  shifted further out by a small amount  from Eq.~(\ref{eq:modradius}).

As observed in \cite{guendel1}, the geometry~(\ref{HHog}) can be viewed as a spherically symmetric collection of straight, rigid, constant-tension, one-dimensional, radially pointing rods (or strings). If $1/\alpha'$ denotes the tension, the total mass of the strings inside a ball of radius $r$ is given by $\;m(r) = 4\pi \int\limits_0^r ~1/\alpha' = 4\pi/\alpha' ~r\;$,  and so the mass function is indeed linear in $r$. Comparing to Eq.~(\ref{mR}), one finds that $\;\alpha'\simeq 8\pi G\; $.
This interpretation of the geometry will also become clear from analyzing the trajectories of infalling objects, which comes up next. We find, to a very good approximation, that objects move on these radial strings in the bulk of the frozen star.

\begin{figure}
\begin{flushleft}  \hspace{-.3in}
\includegraphics[width=0.3\paperwidth]{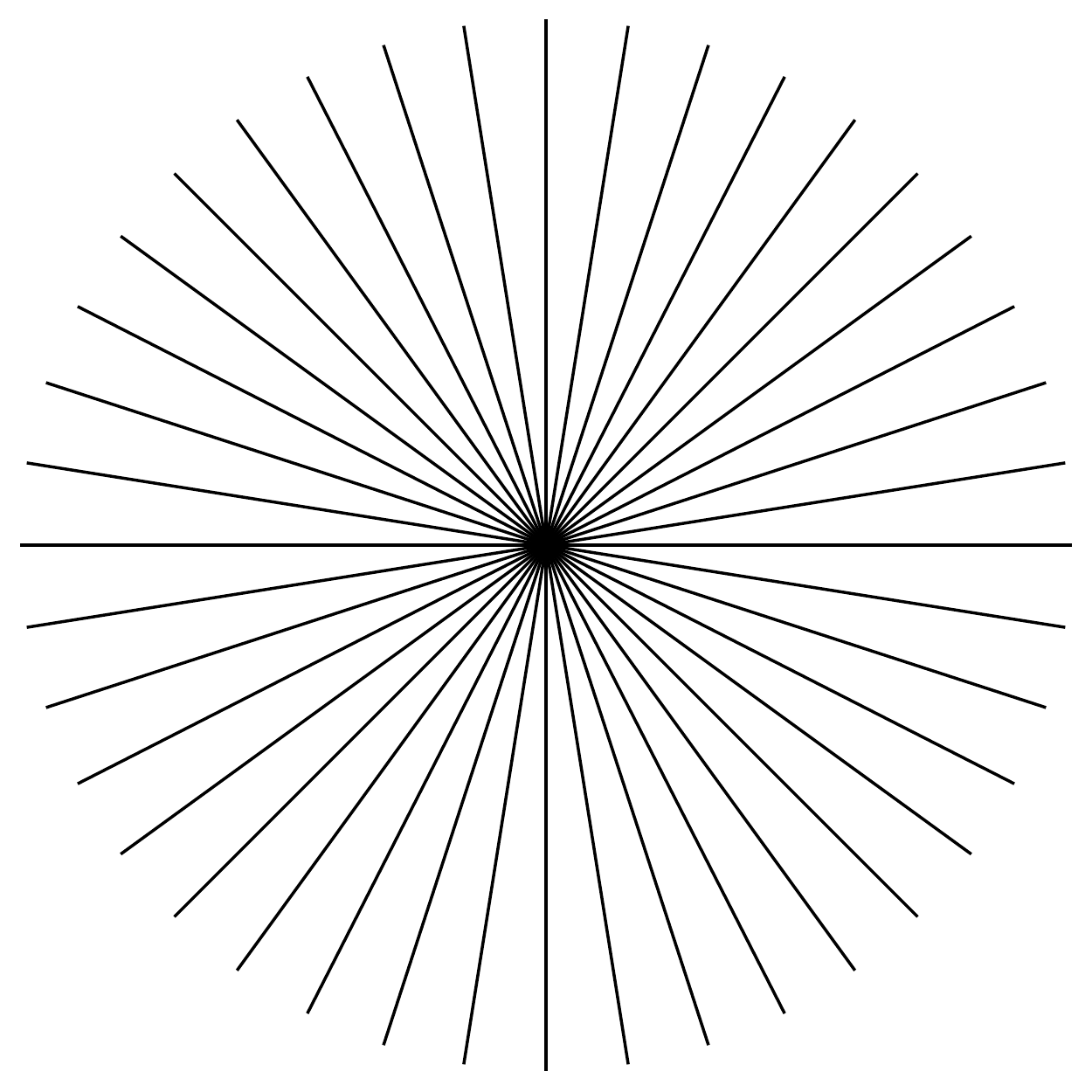},\hspace{0.4in}
\includegraphics[width=0.3\paperwidth]{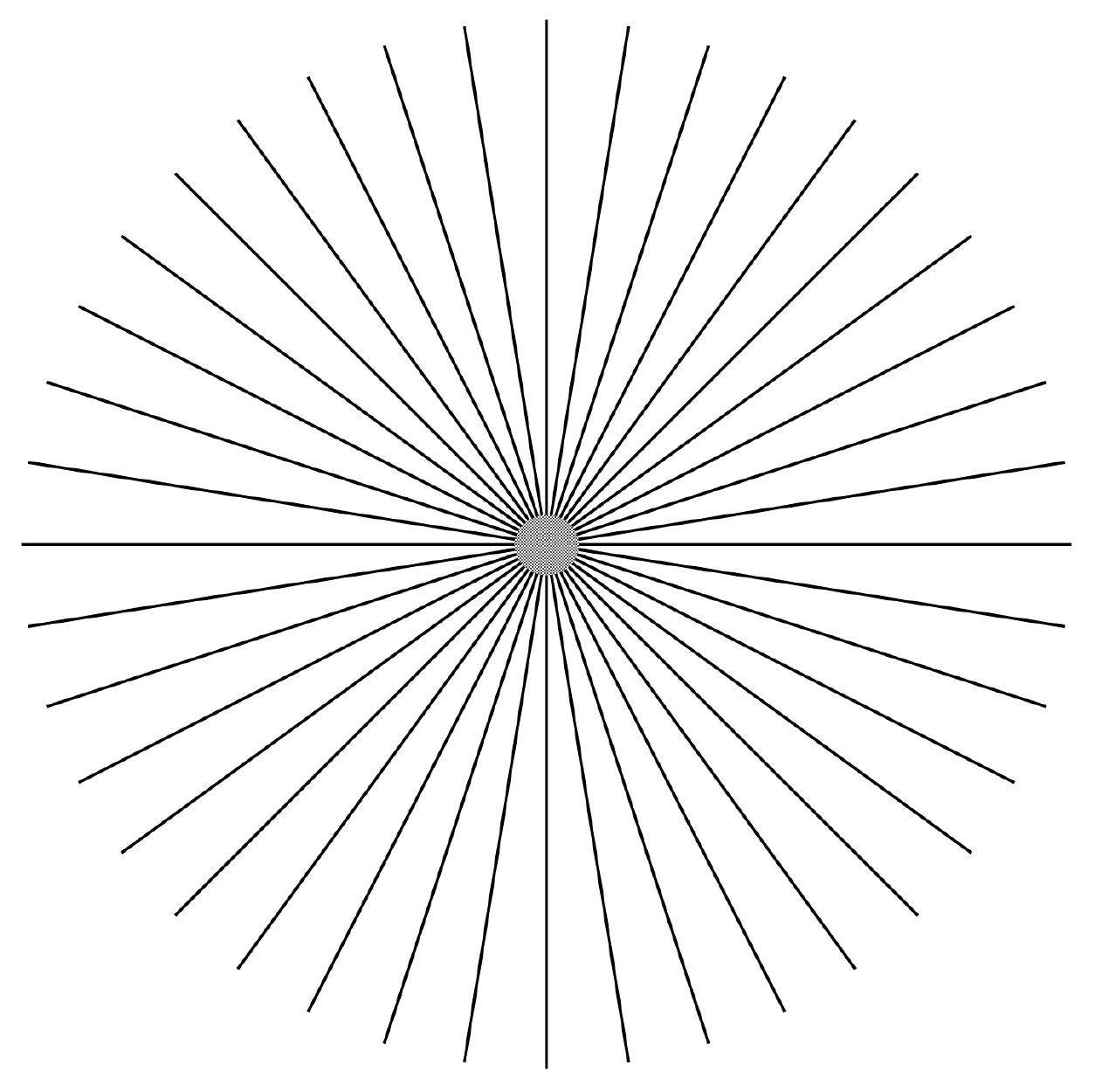}
\par\end{flushleft}
\caption{The frozen star geometry: On the left, unregularized and, on the right, regularized.
\label{fig:HHog}}
\end{figure}

The just-discussed geometry is depicted in Fig.~\ref{fig:HHog}. As is clear from the left panel of the figure, the energy density is formally divergent at the center of the star where all the strings meet. But this divergence is less severe than it looks  because the total mass in this region is small $\;m(r)=r/2\; $. Nevertheless, a solution that smoothes out this divergence is presented in
Section~4 and depicted in the right panel of Fig.~\ref{fig:HHog}. The smoothed-out  solution can be thought of as allowing the strings to bend a little when they reach a certain small distance from the center,  and so they do not all meet at $\;r=0\;$.


\section{The fate of infalling objects}

Here, we are interested  in characterizing the trajectories, both time-like and null, of objects
after falling into the frozen star. We follow the standard textbook discussion to  find the effective gravitational potential that the objects  encounter and use it to understand how they move  through  the star.
Our starting point is the temporal and azimuthal Killing equations.

The temporal Killing equation is
\begin{equation}
f\frac{dt}{d\tau}\;=\;{\mathcal E}\;,\label{eq:tkilling}
\end{equation}
where $\mathcal{E}$, being a conserved quantity,
 is equal to the asymptotic momentum per unit mass.
The azimuthal Killing equation is
\begin{equation}
r^{2}\sin^{2}\theta\frac{d\phi}{d\tau}\;=\;L\;,
\label{killaz}
\end{equation}
such that $L$, also a conserved quantity, is the asymptotic angular momentum per unit mass.


\begin{figure}[t]
\vspace{-2in}
\begin{flushleft}
\includegraphics[width=0.7\paperwidth]{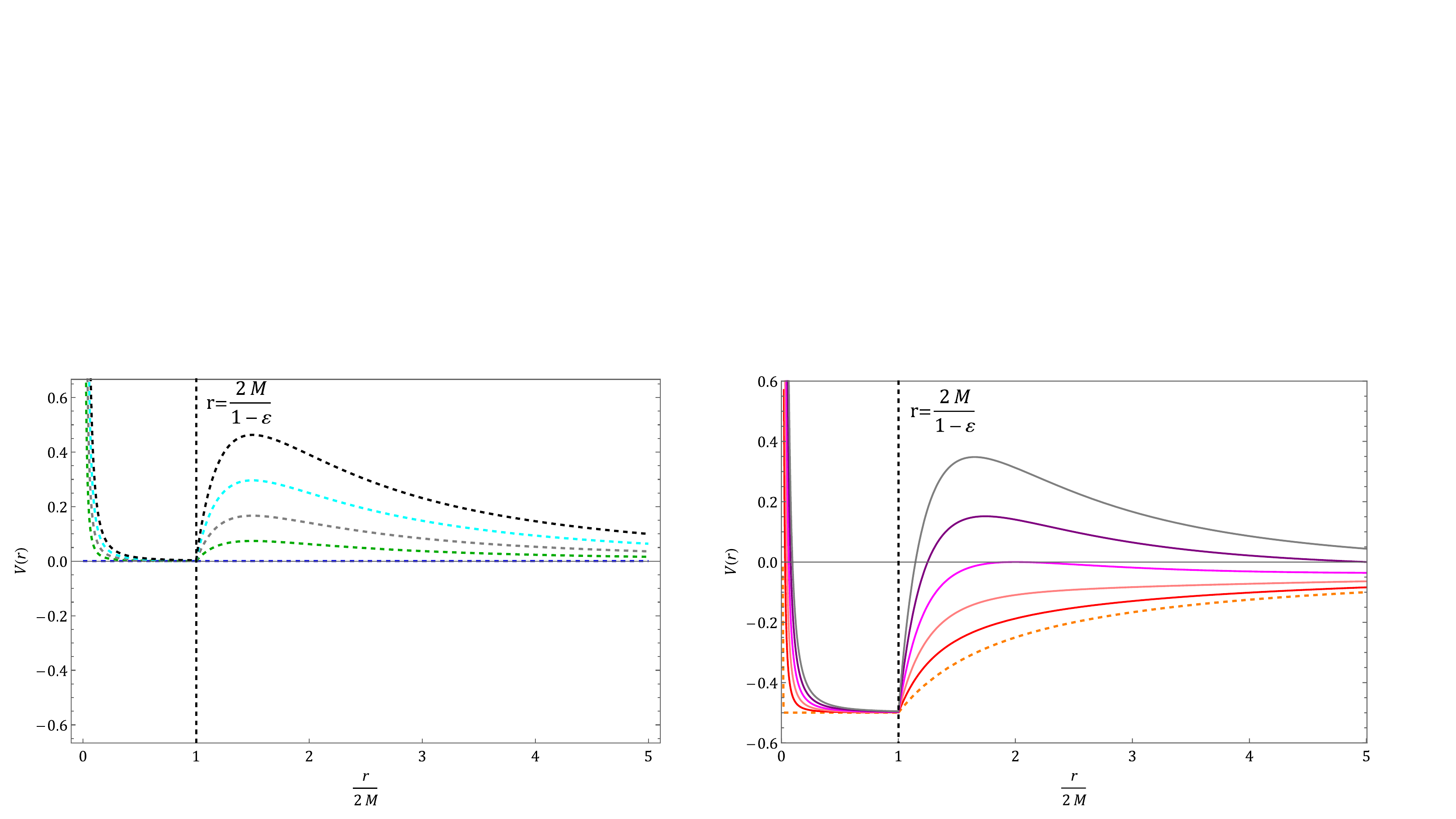}
\par\end{flushleft}
\caption{The gravitational potential of the frozen star for various values of  asymptotic angular
momenta per unit mass.  The potentials
are plotted for both null (left panel) and timelike trajectories (right panel).
\label{fig:gravitationalpotential}}
\end{figure}

One can use the Killing equations to rewrite the velocity-normalization equation for both light ($k=0$)
 and matter ($k=1$),
\begin{equation}
-k\;=\;-f\left(u^{t}\right)^{2}+f^{-1}\left(u^{r}\right)^{2}+r^{2}\left(u^{\Omega}\right)^{2}\;,
\end{equation}
where $u^{i}$ is the 4-velocity component related to the $i$th coordinate. For null trajectories ($k=0$), $u^i$ denotes $dx^i/d\lambda$,  with $\lambda$ being an affine parameter along the trajectory.
As in the Schwarzschild geometry, angular-momentum conservation allows one to consider purely equatorial trajectories.

For an equatorial trajectory, one can rearrange the previous expression into
\begin{align}
\left(u^{r}\right)^{2} & \;=\;f\left[f\left(u^{t}\right)^{2}-r^{2}\left(u^{\Omega}\right)^{2}-k\right]\nonumber \\
 & \;=\;\mathcal{E}^{2}-f\left(\frac{L^{2}}{r^{2}}+k\right)\;
\label{radvel}
\end{align}
and introduce the gravitational potential,
\begin{equation}
\frac{1}{2}\mathcal{E}^{2}\;=\;\frac{1}{2}\left(u^{r}\right)^{2}+V\left(r\right)\;,
\label{EC}
\end{equation}
with
\begin{equation}
V\left(r\right)\;=\;\frac{1}{2}f\left(\frac{L^{2}}{r^{2}}+k\right)\;.
\end{equation}
This potential is depicted in Fig.~(\ref{fig:gravitationalpotential}).

We first discuss in a qualitative manner the properties of trajectories in the frozen star spacetime and then consider  some specific examples.

The trajectories outside  a frozen star of mass $M$ are, by design,  identical to the trajectories in a Schwarzschild  geometry of the same mass. Meaning that trajectories with a large impact parameter do not enter the star, a photon sphere exists at $\;R = 3M\;$ and the redshift factor increases towards the surface of the star.

Trajectories entering the star are of course different but, as argued shortly, the difference is quite subtle from
the perspective of an outside observer.  Inside the star, except for very close to $\;r=0\;$, the potential scales as $\;V\sim f \simeq\varepsilon\ll 1\;$. It follows that the interior trajectories, whether the null or timelike case,  are almost purely radial.  The potential only becomes significant near the center of the star.

In both cases, the 3-velocites are suppressed inside the star by a factor
of $\;\varepsilon\ll 1\;$. It follows that, to a very good approximation, the light-crossing time is
$\;\Delta t \simeq 2R/\varepsilon\;$ and so extremely long. Hence, the  asymptotic coordinate time for any object to move through the star is parametrically much larger than the both the corresponding proper time and the Schwarzschild light-crossing time $4MG/c^3$. To get an idea, the light-crossing time of a frozen star of solar mass, in the case that $\;\varepsilon \sim ~ \frac{1}{100} l_P/R_S\;$ ($l_P$ being the Planck length) is $\;\Delta t\sim 10^{32}\;$~seconds, much larger than the age of the Universe!

In fact, if one also includes quantum-mechanical considerations,  the light-crossing time  greatly exceeds the so-called scrambling time of $R\ln{S_{BH}}$ ($S_{BH}$
is the Bekenstein--Hawking entropy) \cite{scram1,scram2}.  From an external observer's perspective,
this is the time it takes for the object to lose its identity, as it becomes chaotically mixed
with the other degrees of freedom inside the BH. Meaning that, from this perspective and for all practical purposes, an infalling object will adhere onto the surface of the star and never really fall in.

We now turn to discuss some explicit examples, starting with radial infall.
For light ($\;\mathcal{E}=1$, $k=0$), one obtains for the 4-velocity $u$ and 3-velocity $v$, respectively,
\bea
u^{r}\left(r<R\right) &=& \frac{dr}{d\tau}\;=1\;,
\nonumber \\
v^{r}\left(r<R\right)&=&\frac{dr}{dt} \;=\;1-\upsilon^{2} \;.
\eea

For a radially infalling, massive object that is
asymptotically at rest at $\;r\rightarrow\infty\;$ ({\em i.e.}, those with $\;\mathcal{E}=1$),
one obtains for the 4-velocity and 3-velocity, respectively,
\bea
u^{r}\left(r<R\right) &=& \frac{dr}{d\tau}\;=\;\sqrt{1-f}\;=\;\left|\upsilon\right|\;,
\label{v4mass} \\
v^{r}\left(r<R\right)&=&\frac{dr}{dt} \;=\;\left(1-\upsilon^{2}\right)\left|\upsilon\right| \;.
\label{v3mass}
\eea
On the other hand, for massive objects that do have   some initial (asymptotic) radial
3-velocity $\;\upsilon_{0}\;$, the corresponding results are
\begin{align}
u^{r}\left(r<R\right)\;=\;\frac{dr}{d\tau} & \;=\;\gamma\sqrt{\upsilon_{0}^{2}+\left(1-\upsilon_{0}^{2}\right)\upsilon^{2}}\;,
\label{v4massv}
\end{align}
\begin{align}
v^{r}\left(r<R\right)\;=\;\frac{dr}{dt} & \;=\;\left(1-\upsilon^{2}\right)\sqrt{\upsilon_{0}^{2}+\left(1-\upsilon_{0}^{2}\right)\upsilon^{2}}\;,
\label{v3massv}
\end{align}
where $\gamma$ is the corresponding Lorentz factor. One can readily verify that the radial 3-velocity $v^{r}$ cannot exceed the speed of light for any initial $\upsilon_{0}$. Note that Eq.~(\ref{v4massv}) follows from rewriting the energy-conservation equation~(\ref{EC}) at infinity as  $\;\frac{1}{2}\mathcal{E}^{2}=\frac{1}{2}\left(\upsilon_{0}\mathcal{E}\right)^{2}+\frac{1}{2}\;$, solving it for $\mathcal{E}$  and then substituting into Eq.~(\ref{radvel}) with $\;L=0\;$ and $\;k=1\;$. Equation (\ref{v3massv}) requires  the standard conversion from proper to coordinate time.

The crossing time can be calculated for the two types of radial trajectories, but they both lead to
approximately the same outcome. In terms of proper time,
\begin{equation}
\Delta\tau\;=\;\frac{2\left(2M/\upsilon^{2}\right)}{\sqrt{\mathcal{E}^{2}-kf}}
\;=\;\frac{2R}{\sqrt{\mathcal{E}^{2}-\varepsilon k}}\;\simeq \;\frac{2R}{\mathcal{E}}\;,
\end{equation}
and, in terms of asymptotic coordinate time,
\begin{equation}
\Delta t\;=\;\frac{4 M\mathcal{E}} {\upsilon^{2}\left(1-\upsilon^{2}\right)\sqrt{\mathcal{E}^{2}-k\left(1-\upsilon^{2}\right)}} \;=\;\frac{2R}{\varepsilon}\frac{1}{\sqrt{1-\varepsilon k/\mathcal{E}^{2}}}\;\simeq\; \frac{2R}{\varepsilon}\;.
\end{equation}

We now discuss non-radial trajectories.  Because of the $\varepsilon$-suppression of the $L^2$ term in Eq.~(\ref{radvel}), all the trajectories are practically radial for any object until it reaches the proximity of its turning point at a parametrically small distance from the origin.
The discussion is therefore, to some extent, moot because,  after the frozen star has formed, light and  matter will, for all practical purposes, never reach anywhere close to the center. The discussion may still  have some relevance to the trajectories of matter and/or  light that are already trapped inside the frozen star when  it is formed and are extremely close to the center.

All objects with non-zero angular momentum (no matter how little) will be deflected by the centrally located potential barrier. The turning point  $r_{\scriptscriptstyle TP}$ for a given choice of conserved quantities can be
worked out  by setting the left-hand side of Eq.~(\ref{radvel}) to zero,
\be
r_{\scriptscriptstyle TP}\;=\; L\sqrt{\frac{f}{\mathcal{E}^2-kf}}\;.
\label{TeePee}
\ee
Moreover, since $\;f=\varepsilon \ll 1\;$ and $\;\mathcal{E} \gtrsim 1\;$,
\be
r_{\scriptscriptstyle TP}\;\approx\; \sqrt{\varepsilon} \frac{L}{\mathcal{E}}\;,
\label{TeePee2}
\ee
where  $|L/\mathcal{E}|$ is identifiable as the impact parameter of the trajectory.


\begin{figure}[t]
\begin{center}
\vspace{-1.5in}
\includegraphics[width=0.35\paperwidth]{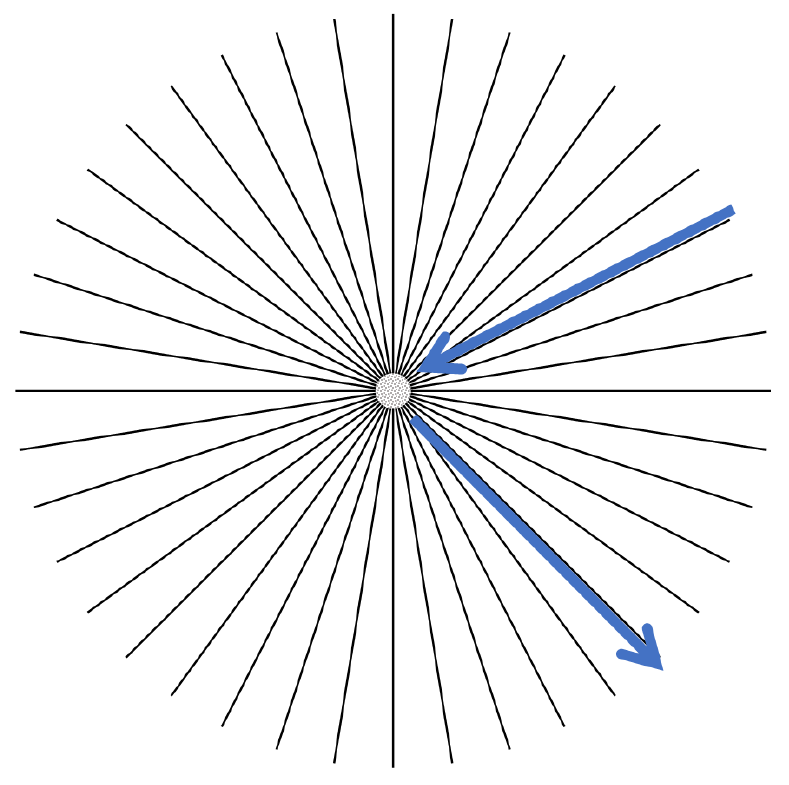}
\par\end{center}
\caption{Trajectory deflecting from the frozen star geometry. As in the Schwarzschild geometry, trajectories become radial as they approach the surface of the frozen star. They stay almost radial until they reach the core of the star and are then deflected into another almost-radial trajectory.
\label{fig:Penrose}}
\end{figure}

It is  interesting to ask about the deflection angle in the current scenario. To address
this, we start by using  Eq.~(\ref{killaz}) to solve for $\frac{d\phi}{d\tau}$ and  Eq.~(\ref{radvel}) to solve for $\;u^r=\frac{dr}{d\tau}\;$. We then divide
the former by the latter.

Let us specifically discuss  light deflection, so that  $\;\mathcal{E}=1\;$ and $\;k=0\;$. In this case,
\be
r_{\scriptscriptstyle TP}\;=\; \sqrt{\varepsilon} L\;,
\label{TeePee3}
\ee
and then, on the equatorial plane,
\be
\frac{d\phi}{dr}\;=\;\frac{L}{r^2\sqrt{1 -\frac{r_{\scriptscriptstyle TP}^2}{r^2}}}\;.
\ee
The total angular deflection inside of  the star is given by the integral
\bea
\Delta \phi &=& 2 L\int^R_{r_{TP}} \frac{dr}{r^2\sqrt{1 -\frac{r_{\scriptscriptstyle TP}^2}{r^2}}}\;.
\eea
In terms of the dimensionless variable $\;x=\dfrac{r}{r_{\scriptscriptstyle TP}}\;$,
this becomes
\be
\Delta \phi\;=\; \frac{1}{\sqrt{\varepsilon}}\int_{1}^{R/r_{\scriptscriptstyle TP}} \frac{dx}{x\sqrt{x^2-1}}\;,
\ee
with the result that
\be
\Delta \phi\;=\; \frac{\pi}{2} \frac{1}{\sqrt{\varepsilon}}+\dots\;,
\ee
where relative corrections of order $r_{\scriptscriptstyle TP}^{ }/R\ll 1$ were neglected. The resulting angular deflection is very large and  mostly accumulated near the turning point of the trajectory. The photon  spins around the core of the BH and emerges onto  another  almost-radial trajectory.

\section{Smoothing the core}

We have now seen that the frozen star metric can be made regular throughout the spacetime while maintaining most of the essential features of the model.  Nevertheless, even with $\;f=|g_{tt}|=g^{rr}>0\;$, the density profile for the interior is $\rho\sim \frac{1}{r^2}$, which diverges at $\;r=0\;$. The divergence is rather mild, as the mass in the core region is parametrically small. Nevertheless, our objective is to complete the frozen star model such that it is regular everywhere, and we accomplish this  by applying a suitable regularization scheme at small values of $r$.
General considerations suggest that the regularization scale is some fundamental scale such as the string length, which is  small but larger than the Planck length.

Our scheme will entail  examining a sphere of radius $2\eta$ centered around $\;r=0\;$, where
$\;\eta\ll R\;$. The sphere consists of an inner sphere of radius $\;\eta\;$, surrounded by a spherical shell of width $\;\eta\;$. The metric function $f(r)$  of the shell connects smoothly at $\;r=2\eta\;$  to that  of the bulk of the frozen star,  $\;f=\varepsilon\;$,  {\em and} smoothly at $\;r=\eta\;$ to that of the inner sphere, which will be discussed later.  The regularization prescription is not unique. We present here the simplest procedure that works, allowing us to verify that such a regularization does not change the physics of the frozen star and that it  is consistent with all the fundamental constraints, such as the positivity of $\rho$, the integrity  of the  null energy condition and so on.


\begin{figure}[t]
\begin{center}
\vspace{-1.5in}
\includegraphics[width=0.35\paperwidth]{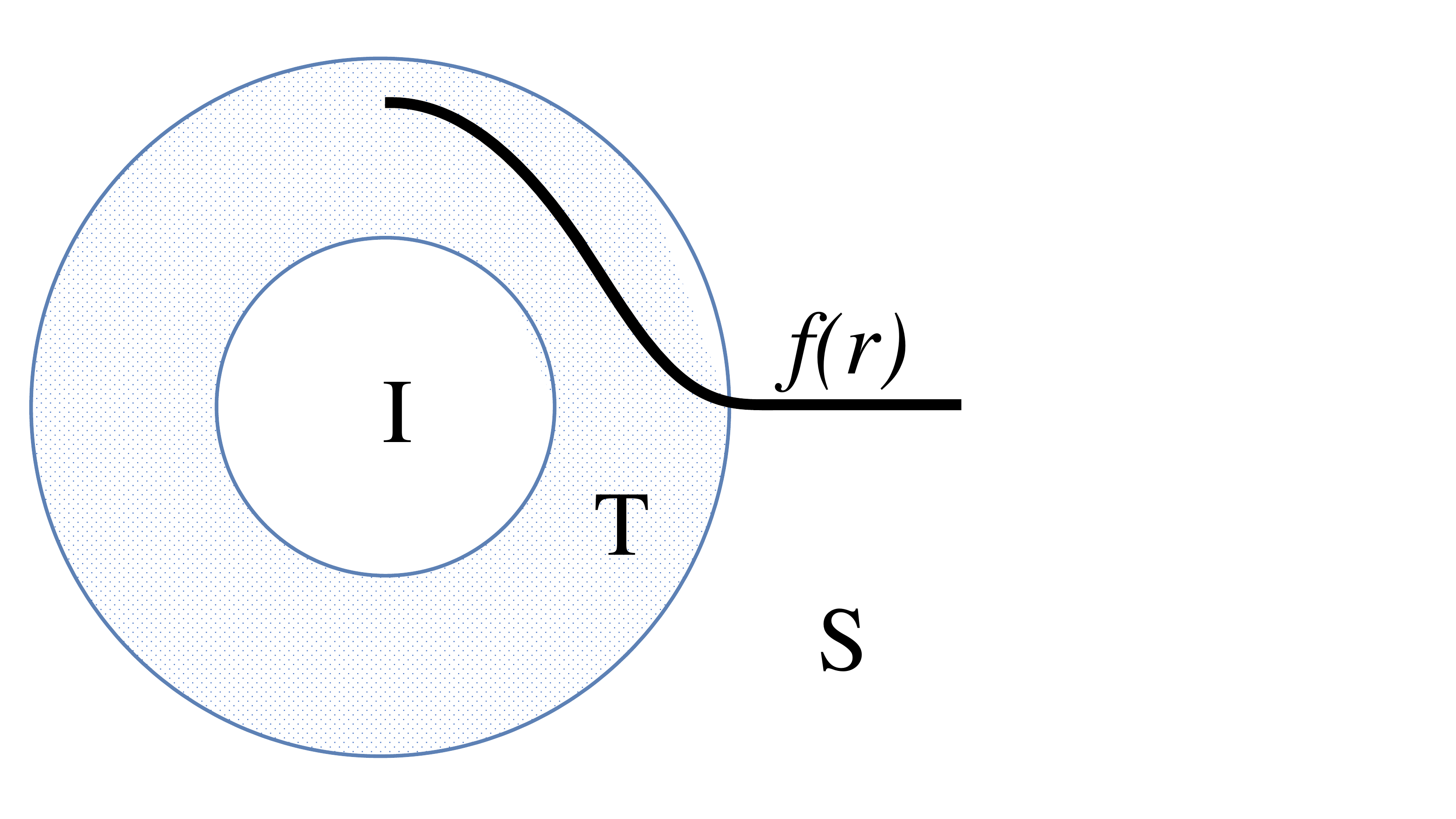}
\par\end{center}
\caption{Smoothed core of the frozen star. The interior of the core $r\le \eta$  is denoted by $I$, the transitional layer $\eta\le r\le 2\eta$ by $T$ and the frozen star bulk $r\ge 2 \eta$  by $S$. The metric function $f(r)$ is depicted by the thick black line. The value of $f$ at the center is $f(0)=1$ and it decreases to $\varepsilon$ at $r=2 \eta$.
\label{fig:smooth}}
\end{figure}

We will assume the same form of  line element as presented in Section~2,
\be
ds^2\;=\;-f(r)dt^2+\frac{1}{f(r)}dr^2+r^2(d\theta^2+\sin^2\theta d\phi^2)\;,
\label{line}
\ee
along with same form of stress  tensor, $\;T^\mu_{\;\;\nu}= \text{diag}
(-\rho, -\rho, p_\perp, p_\perp)\;$. Also as in Section~2, $\;f=\varepsilon\ll 1\;$
is the metric function for the star interior, extending  from  the outer surface at about $\;R=2M(1+\varepsilon)\;$ down to the surface $\;r=2\eta\;$.

Let us write the metric function for the region of interest, $\;0<r<2\eta\;$, as
\be
f(B,\eta,r)\;=\;
\begin{cases}
	f_I(B,\eta,r)\;,\quad  r<\eta\;,\\
	f_T(B,\eta,r)\;, \quad \eta<r<2\eta\;,\\
	f_S\;,\quad r>2\eta\;,
\end{cases}\label{FBR2}
\ee
where $f_S$ is the metric function of the frozen star interior, $f_T$ is the metric function of the translational layer,  $f_I$ is the metric function  within  the inner sphere of radius $\eta$ and  $B$ is the energy density at
the center or $\;\rho(r=0)=B\;$. The constant $B$  then has  dimensions of
inverse length squared and is  assumed to be positive. As shown in the Appendix,
the self-consistency of our model requires the  dimensionless combination $B\eta^2$ to be a number of order unity, making it  convenient to redefine $\;B\eta^2\to B\;$.

In what follows, we will simplify notation by working in units with  $\;8\pi G=1\;$  and all other dimensional
quantities will be in units of $R$. Equivalently, we are setting  $\;R=1\;$ and rescaling other quantities appropriately,
so that the parameter $\eta$ is a
small constant number and $r$ is similarly small in the region of interest.

It is useful to define the energy density and transverse pressure in the same way as  done in Eq.~(\ref{FBR2}),
\be
\rho(B,\eta,r)\;=\;
\begin{cases}
	\rho_I(B,\eta,r)\;,\quad  r<\eta\;,\\
	\rho_T(B,\eta,r)\;, \quad \eta<r<2\eta\;,\\
	\rho_S\;,\quad r>2\eta\;,
\end{cases}\label{DenBR2}
\ee
\be
p_\perp(B,\eta,r)\;=\;
\begin{cases}
	p_{I\perp}(B,\eta,r)\;,\quad  r<\eta\;,\\
	p_{T\perp}(B,\eta,r)\;, \quad \eta<r<2\eta\;,\\
	p_{S\perp}\;,\quad r>2\eta\;.
\end{cases}\label{PBR11}
\ee
We can then derive $f_I$ by solving the Einstein equations and, in turn, $f_T(B,\eta,r)$ by matching smoothly to $f_I$ and to $f_S$ at their respective connecting surfaces.

\subsection{Matching conditions and the function $f_I$}

Continuing  with our (non-unique) regularization procedure, we require  that $\rho$ decreases monotonically  in the region $\;0<r<{\eta}\;$
from its maximal value of $B$ at the center. Let us then consider the function $\;\rho_I({B},{\eta},r)=-r+\frac{{B}}{{\eta}^2}\;$. Using this form and  Einstein's equations, we can derive the metric function and its first two derivatives,
\be
f_I({B},{\eta},r)\;=\;1-\frac{r^2}{3}\frac{{B}}{{\eta}^2}+\frac{r^3}{4}\;,\label{fI}
\ee
\be
f_I'({B},{\eta},r)\;=\;-\frac{2r}{3}\frac{{B}}{{\eta}^2}+\frac{3r^2}{4}\;,
\ee
\be
f_I''({B},{\eta},r)\;=\;-\frac{2}{3}\frac{{B}}{{\eta}^2}+\frac{6 r}{4}\;.
\ee
The last term in each of these equations is subleading,  but we need to retain them  to ensure the continuity of $f$ and of the energy density.

Equations~(\ref{FBR2}), (\ref{DenBR2}) and (\ref{PBR11}) can now be rewritten as follows:
\be
f({B},{\eta},r)\;=\;
\begin{cases}
	1-\dfrac{r^2}{3}\dfrac{{B}}{{\eta}^2}+\dfrac{r^3}{4}\;,\quad  r<{\eta}\;,\\
	f_T({B},{\eta},r)\;, \quad {\eta}<r<2 {\eta}\;,	\\
	\varepsilon\;,\quad r>2 {\eta}\;,
\end{cases}
\label{FBR3}
\ee
\be
 \rho({B},{\eta},r)\;=\;
\begin{cases}
	-r+\dfrac{{B}}{{\eta}^2}\;,\quad  r<{\eta}\;,\\
	\rho_T({B},{\eta},r)\;, \quad {\eta}<r<2 {\eta}\;,\\
	\dfrac{1-\varepsilon}{r^2}\;,\quad r>2{\eta}\;,
\end{cases}\label{DenBR1}
\ee
\be
p_\perp({B},{\eta},r)\;=\;
\begin{cases}
	\dfrac{3r}{2}-\dfrac{{B}}{{\eta}^2}\;,\quad  r<{\eta}\;,\\
	p_{T\perp}({B}, {\eta},r)\;, \quad{\eta}<r<2 {\eta}\;,\\
	0\;,\quad r>2 {\eta}\;.
\end{cases}\label{PBR}
\ee
It should be kept in mind that the validity of our model depends on  ${B}$ being the order of unity. In fact, the allowable
range of $B$ values happens to be   $\frac{6}{5}< B\le \frac{30}{19}$ .

The form of the metric $f_T({B},{\eta},r)$, is found by adopting a polynomial ansatz for it  in terms of $(r-{\eta})$. The order of the polynomial and, thus, the number of adjustable parameters in the ansatz is determined by the number of relevant matching conditions. Here, we are requiring
that $f({B},{\eta},r)$ and its first two derivatives be continuous at both ends of the interpolating  layer, $\;r=\eta\;$ and $\;r=2\eta\;$, which necessitates  a fifth-order polynomial. These matching conditions and  the restriction of $B$
as discussed  is enough to ensure that the energy density, pressure and their first and second derivatives are all continuous. Additionally, these conditions guarantee that $f({B},{\eta},r)$ is positive,  $\rho$ is positive and $\rho'$ is negative. Interestingly, they also result in a negative value of $p_{\perp}$; however, the null energy condition is
not violated.

The details of the matching procedure, the results and the detailed discussion of the constraints have been relegated to the Appendix.

\section{Causal structure}

For any finite $\varepsilon$, the  Penrose coordinates for a frozen star can be defined in the standard way. One starts with the coordinates
\bea
X(r,t)\;=\; -e^{-\frac{1}{2}[t-r_*(r)]}\;, \\
Y(r,t)\;=\; -e^{-\frac{1}{2}[t+r_*(r)]}\;,
\eea
where $r_*$ is the usual tortoise coordinate,
\be
dr_*\;=\; \frac{dr}{f(r)}\;.
\ee
For $\;r>R\;$, this is the standard Schwarzschild form, $\;r_*= r+R_S\ln(r-R_S)\;$.
However, for $\;r \le R\;$  and so inside of  the frozen star,
\be
r_*\;=\;\frac{r}{\varepsilon}\;.
\ee

In the transitional region near the surface of the frozen star and in the smoothed central region, the functional form of $r_*$ changes, but the numerical values do not differ much. For the purpose of the discussion of the causal structure, we can ignore these differences.

The Penrose coordinates can now be defined as
\bea
U(r,t)\;=\;\arctan(X(r,t))\;, \\
V(r,t)\;= \;\arctan(Y(r,t))\;.
\eea

The Penrose diagram of the frozen star is depicted in Fig.~\ref{fig:Penrose}. For any finite $\varepsilon$, it is similar to that of the whole of Minkowski space.  The thick blue line marks the position of the surface of the star. For small enough $\varepsilon$,  the whole interior of the star is null.


\begin{figure}
\begin{center}
\includegraphics[width=0.60\paperwidth,angle=0]{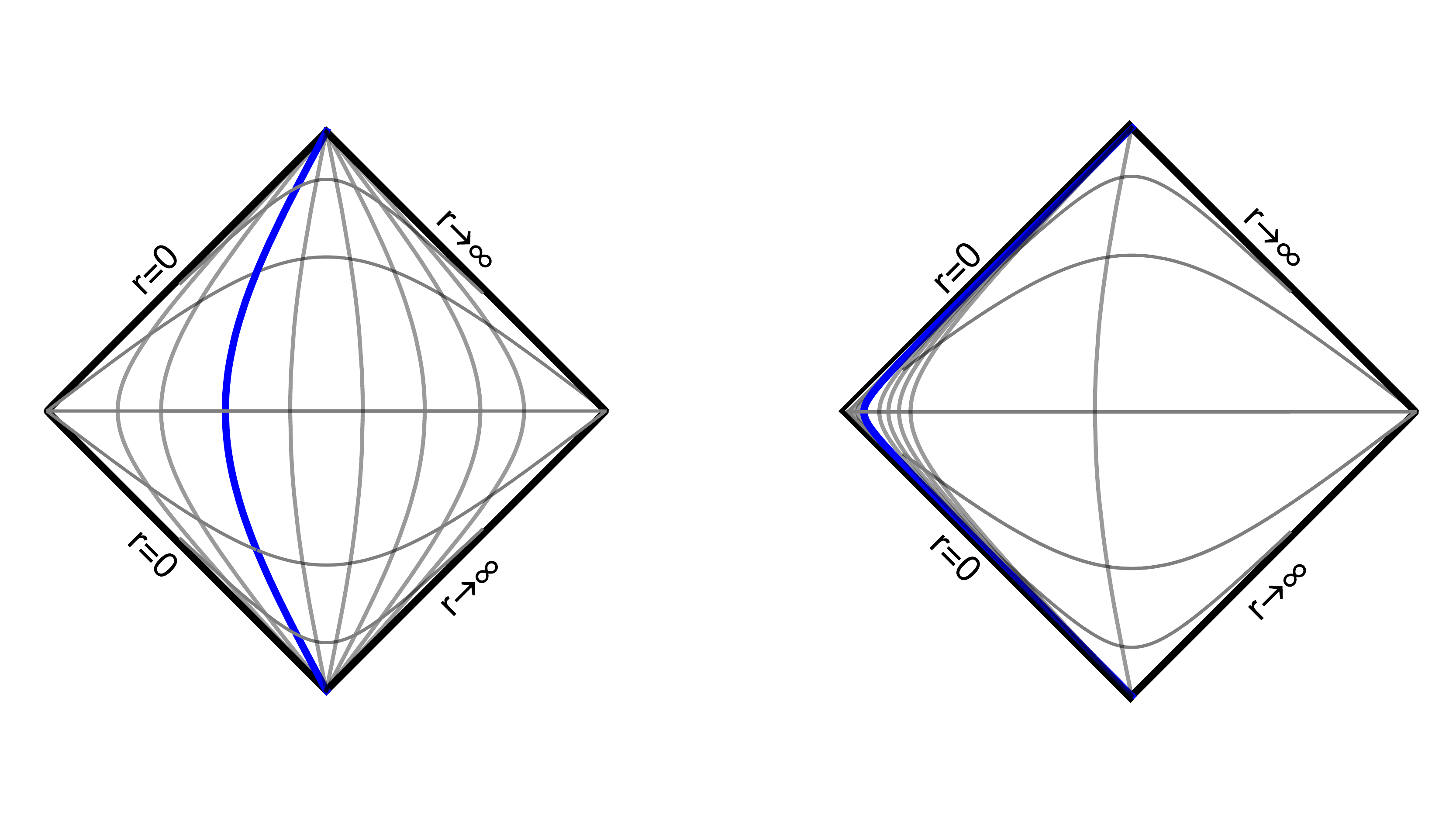}
\par\end{center}
\caption{The Penrose diagram of a frozen star for finite $\varepsilon$ (left) and for $\varepsilon\to 0$ (right). The thick, blue line marks the position of the surface of the star.  The diagram for $\varepsilon=0$ is degenerate: all radial surfaces in the interior of the star collapse to a single null surface.
\label{fig:Penrose}}
\end{figure}

The causal structure at $\;\varepsilon=0\;$ is more subtle. To help visualize this rather unorthodox geometry, it is useful to recast the discussion in terms of a set of adapted Kruskal-like coordinates. To do this, we introduce  the coordinate transformation
\begin{eqnarray}
U_0 & \;=\;\exp\left[-\frac{r}{\sqrt{\varepsilon}}+\sqrt{\varepsilon}~ t\right]\;,\\
V_0 & \;=\;\exp\left[-\frac{r}{\sqrt{\varepsilon}}-\sqrt{\varepsilon}~ t\right]\;,
\end{eqnarray}
which results in a Kruskal-like form of  line element for the interior of the frozen star,
\begin{equation}
ds^{2}\;=\; -\frac{dU_0dV_0}{U_0V_0}+r^{2}\left(U_0,V_0\right)d\Omega^{2}\;.
\label{eq:kruskal}
\end{equation}

Radially null light cones ($ds^{2}=0\;$ at constant $\theta,\;\phi$) appear
as horizontal and vertical lines of constant $U_0,\;V_0$.
The following ratio,
\begin{equation}
\frac{U_0}{V_0}\;=\;\exp\left[2\sqrt{\varepsilon}~t\right]\;,
\end{equation}
reveals that trajectories  of constant $t$  appear in the form of straight
lines. Similarly, the product $U_0,\;V_0$ shows that lines
of constant  $r$ have the form
\begin{equation}
U_0 V_0\;=\;\exp\left[-\frac{2r}{\sqrt{\varepsilon}}\right]\;. \label{eq:TRinr}
\end{equation}

In the  limit  $\varepsilon \rightarrow 0\;$, the product $U_0V_0$ vanishes unless $\;r\lesssim \sqrt{\varepsilon}\;$ and the ratio $U_0/V_0$ approaches $1$ unless $\;t \gtrsim 1/\sqrt{\varepsilon}\;$. it follows that when $\varepsilon=0\;$, the entire interior geometry is defined by the lines $\;U_0=0\;$ and $\;V_0=0\;$. This implies, in turn, that the $\;\varepsilon\rightarrow 0\;$ limiting case is effectively 1+1 dimensional as each 3-sphere collapses to a point. This is interesting in that the polymer model has the effective thermodynamics of a 1+1-dimensional
radiative matter system.

\section{Briefly on stability}

One can repeat the analysis on stability in \cite{popstar} to show that a choice of $\;f=1-v^2\;$
different from zero has no bearing on the ultra-stability of the frozen star, nor does the inclusion
of the regularized core as discussed in Section~4 and the Appendix. As it turns out, the exact value of $f$ has essentially no role in stabilizing the star against perturbations; this really falls under the purview of the
equation-of-state  condition $\;\rho+p_r=0\;$. In fact, repeating the analysis, one finds that the only difference between a vanishing $f$ and a constant $f$ is that the perturbation of $g_{tt}$ --- what we called $H_0$ in \cite{popstar} ---
need not be set to  zero {\em a priori}; it could just as well  be set to a non-zero constant. However, any choice but zero would violate the equation of state over macroscopic time scales, meaning that zero is the uniquely correct choice after all. In short, the thawing out of a frozen star requires deviations from   $\;\rho+p_r=0\;$ and thus from $\;{\widetilde f}=f\;$.

As for the geometry of the regularized core, this can be incorporated into our stability  analysis just as the was done for the transitional layer at the outermost surface.

\subsection*{Overview}

We have furthered our investigations into a classical but regular model for a BH interior that is known
as the frozen star. This classical version of our highly quantum polymer model has two prominent features:
maximally negative pressure throughout and each spherical slice of the interior is a surface of infinite redshift. Here, we have relaxed the latter (but not the former) on the basis that the idealized situation
is probably not physically realistic and that the metric is now regular throughout the spacetime.
We used this softened picture to  understand the fate of infalling matter and found that a particle
with any amount of angular momentum will be reflected before reaching the center of the star.  We described a regularization procedure at the core of the frozen star resulting in a completely regular metric. The technical details of the regularization procedure were presented in the Appendix. Meanwhile, Kruskal-like coordinates were introduced so as to provide us with a metric that remains regular even when
the  limiting case of slice-by-slice infinite redshift is restored.

Now having a regular and classical metric to work with, we are well positioned to compare and contrast our model with that of a general-relativistic BH. This should soon be possible from the analysis of gravitational  waves  emerging from BH mergers, as out-of-equilibrium physics is the key to understanding how
the ``true'' theory of gravity may distinguish itself from Einstein's. For further reading on this perspective, see \cite{ridethewave,spinny,collision,QLove,CLove}.

\section*{Acknowledgments}
We thank Eran Palti for pointing out the necessity of smoothing the core of the frozen star.
The research is supported by the German Research Foundation through a German-Israeli Project Cooperation (DIP) grant ``Holography and the Swampland.''
The research of AJMM received support from an NRF Evaluation and Rating
Grant 119411 and a Rhodes  Discretionary Grant SD07/2022.  AJMM thanks Ben Gurion University for their hospitality during his visit.

\newpage

\begin{appendices}

\section{Regularizing  the core}

Here, we  present the details of the calculation for determining $f_T$, the metric function for
the translational  layer in the core,   and the allowed  range of values for   $\;B=\rho(r=0)\;$.
One should have already read Section~4 before proceeding any further.

As discussed in the main text, the continuity of the metric and its first
two derivatives require six matching conditions for $f_T$ (three at each of the two connecting surfaces).
These conditions are
\be
f_T({B},{\eta}, r=2{\eta})\;=\;\varepsilon\;,
\label{FirstMat}
\ee
\be
f_T({B},{\eta},r={\eta})\;=\; f_I({B},r={\eta})\;,
\ee
\be
f_T'({B},{\eta},r=2{\eta})\;=\;0\;,
\ee
\be
f_T'({B},{\eta},r={\eta})\;=\; f_I'({B},r={\eta})\;,
\ee
\be
f_T''({B},{\eta},r=2{\eta})\;=\;0\;,
\ee
\be
f_T''({B},{\eta},r={\eta})\;=\;f_I''({B},r={\eta})\;,\label{SixMat}
\ee
where $f_I$ is the metric function for the innermost region of the core.
Conditions~(\ref{FirstMat})-(\ref{SixMat}) also ensure the continuity of the  energy density, its first derivative and the transverse pressure. (It should be kept in mind that the radial pressure follows automatically as the negative of   $\rho$.)

Assuming that $f_T({B},{\eta},r)$ adopts the polynomial form  (the alphabetically ordered letters are yet-to-be-determined
coefficients)
\be
f_T({B},\eta,r)\;=\;a+b(r-{\eta})+c(r-{\eta})^2+d(r-{\eta})^3+e(r-{\eta})^4+g(r-{\eta})^5 \label{Polynom}
\ee
and using the matching conditions~(\ref{FirstMat})-(\ref{SixMat}), one  can find the coefficients of the polynomial and,  thus,  also derive $\rho_T$ and $p_{T\perp}$ by way of Einstein's equations.


\subsection{Results of $f$, $\rho$ and $p_\perp$ as a function of $B, \eta, r$}

In this section of the Appendix, we present the expressions for the metric function, energy density and transverse pressure, respectively.  Let us recall the conventions; namely that $\;8\pi G =1\;$,
$B$ really means the dimensionless product $B\eta^2$  (which happens to be of order unity) and all other dimensional parameters are expressed in
units of $\;R=1\;$

For simplicity, we have neglected subleading terms; meaning  that all of the terms
in a given equation  are of the same order in the small parameters
$\varepsilon$, $\eta$, $r$ and $(r-\eta)$ (the last two being small  specifically in the region of interest). The resulting expressions
go as
\bea
&& f_T(B,\eta,r)\;=\;\frac{1}{12} \left(-4 B+12\right)+\frac{\left(26 B-36\right) (r-\eta )^5}{6
	\eta ^5}+\frac{\left(-34 B+45\right) (r-\eta )^4}{3 \eta ^4}\cr
&& +\frac{\left(100
	B-120\right) (r-\eta )^3}{12 \eta ^3}+\frac{1}{12} \left(-\frac{4
	B}{\eta ^2}\right) (r-\eta )^2+\frac{1}{12} \left(-\frac{8 B}{\eta }\right)
(r-\eta )\;,	\hspace{.3in}
\label{Q}
\eea
\bea
r^2\rho_T(B,\eta,r)&=&\frac{1}{12} \left(4 B-12\right)-\frac{\left(26 B-36\right) (r-\eta )^5}{6
	\eta ^5}-\frac{\left(-34 B+45\right) (r-\eta )^4}{3 \eta ^4}\cr
&-&\frac{\left(100
	B-120\right) (r-\eta )^3}{12 \eta ^3}-\frac{1}{12} \left(-\frac{4
	B}{\eta ^2}\right) (r-\eta )^2-\frac{1}{12} \left(-\frac{8 B}{\eta }\right)
(r-\eta )\cr
&-&r \bigg(\frac{1}{12} \left(-\frac{8 B}{\eta }\right)+\frac{5 \left(26 B-36\right) (r-\eta )^4}{6 \eta ^5}+\frac{4 \left(-34 B+45\right)
	(r-\eta )^3}{3 \eta ^4}\cr
&+&\frac{\left(100 B-120\right) (r-\eta )^2}{4 \eta
	^3}+\frac{1}{6} \left(-\frac{4 B}{\eta ^2}\right) (r-\eta )\bigg)+1
\label{Rho}\;,
\eea
\bea
&& 2r p_{T\perp}(B,\eta,r)\;=\; r \Bigg(\frac{1}{6} \left(-\frac{4 B}{\eta ^2}\right)+\frac{10 \left(26 B-36\right) (r-\eta )^3}{3 \eta ^5}+\frac{4 \left(-34 B+45\right) (r-\eta )^2}{\eta^4}\cr
&+&\frac{\left(100 B-120\right) (r-\eta )}{2 \eta ^3}\Bigg)+2 \bigg(\frac{1}{12}
\left(-\frac{8 B}{\eta }\right)+\frac{5 \left(26 B-36\right) (r-\eta)^4}{6 \eta ^5}\cr
&+&\frac{4 \left(-34 B+45\right) (r-\eta )^3}{3 \eta^4}+\frac{\left(100 B-120\right) (r-\eta )^2} {4\eta^3}+\frac{1}{6} \left(-\frac{4 B}{\eta ^2}\right) (r-\eta )\bigg)\;.
\label{p-bar}
\eea
One can see that these results do not depend on $\varepsilon$ since terms involving it are subleading.

By applying some additional physical constraints, one can find a range of values for $B$, which we do next.

\subsection{Non-negativity of $f_T$}

As  already seen in Section~4, the metric function is non-negative in the innermost region $\;0<r<\eta\;$. We need to verify  that the same is true for the transitional region $\;\eta <r < 2\eta\;$ .

One can simplify Eq.~(\ref{Q}) to obtain
\be
f_T(B,\eta,r)\;=\;\frac{(r-2 \eta )^3 \left(3 \eta ^2 \left(3B-4\right)+r^2 \left(13B-18\right)-3 \eta  r \left(7B-9\right)\right)}{3 \eta ^5}\;.\hspace{.8in}
\label{EsQ}
\ee
The denominator is clearly positive. As for the numerator, the factor $(r-2 \eta )^3$ is negative for $\;\eta<r<2\eta\;$. Thus, in order for $f_T(B,\eta,r)$ to be positive, the other factor in the numerator must then be negative.
This will happen for $\;0<B\leq\frac{30}{19}\;$.


\subsection{Positivity of $\rho_T$}

Simplifying Eq.~(\ref{Rho}), we have
\bea
&& r^2\rho_T(B,\eta,r)\;=\;
\frac{r^5 \left(36-26 B\right)+15 \eta  r^4 \left(11B-15\right)-8 \eta ^4 r
	\left(23B-30\right)}{\eta ^5 r^2}\cr
&&+\frac{\eta ^5 \left(24B-31\right)+10
	\eta ^3 r^2 \left(41B-54\right)+r^3 \left(520 \eta ^2-388 B\eta
	^2\right)}{\eta ^5 r^2}\;.
\label{EsRho}
\eea
The denominator is clearly positive. As for the numerator, it is positive for $\;0<B<2.53\;$.


\subsection{Negativity of $(r^2\rho_T)'$}

Differentiating Eq.~(\ref{EsRho}) with respect to $r$ and then simplifying, one can show that
\bea
	(r^2\rho_T)'(B,\eta,r)&=&\frac{-48B \eta ^5+30 \eta  r^4 \left(11 B-15\right)+r^5 \left(108-78B\right)+62 \eta ^5}{\eta ^5 r^3}\cr
	&+&\frac{r^3 \left(520 \eta ^2-388B\eta ^2\right)+8
		\eta ^4 r \left(23B-30\right)}{\eta ^5 r^3}\;.
\eea
The denominator is positive,  whereas the numerator is negative as long as  $\;\frac{6}{5}<B<1.84\;$ is satisfied.

Combining the results of Sections~A.2,~A.3 and~A.4,  we find that  $B$ is restricted to the advertised range
of values,
\be
\frac{6}{5}\;<\;B\;\leq\;\frac{30}{19}\;.
\label{B-cond}
\ee


\subsection{Limited negativity of $p_\perp$}

The transverse pressure $p_{T\perp}(B,\eta,r)$ has to connect to a negative function at $\;r=\eta\;$ and to $0$ at $\;r=2\eta\;$, necessitating a sign change. However, the transverse form of the  null energy condition, $\;\rho_{T}(B,\eta,r)+p_{T\perp}(B,\eta,r)\geq 0\;$, is never violated  for the  allowable range~(\ref{B-cond}) of $B$ values.


\subsection{Conditions for continuity of first and second derivative of $\rho_T$ and $p_{T\perp }$}

In this section, we will use the exact expressions of the energy density and transverse pressure, since the truncated forms in Eq.~(\ref{Rho}) and Eq. (\ref{p-bar}) will lead to an  apparent discontinuity in their respective derivatives.
To start, the first derivative of the energy density $\rho(B,\eta,r)$ can be expressed as
\be
\rho'(B,\eta,r)\;=\;
\begin{cases}
	-1\;, \quad  r<\eta\;,\\
	\rho_T'(B,\eta,r)\;, \quad \eta<r<2\eta\;, \\
	-\frac{2(1-\varepsilon)}{r^3}\;,\quad r>2\eta\;.
\end{cases}
\label{DenBR'}
\ee
If the energy denisty  is continuous, it has to satisfy the matching conditions $\;\rho_T'(B,\eta,r=\eta)=-1\;$ and $\;\rho_T'(B,\eta,r=2\eta)=-\frac{2(1-\varepsilon)}{(2\eta)^3}\;$, both of which can be readily verified using our previous
results for the metric function and Einstein's equations.

The same basic procedure can be repeated for $\rho''(B,\eta,r)$ and $p_\perp'(B,\eta,r)$,
\be
\rho''(B,\eta,r)\;=\;
\begin{cases}
	0\;,\quad  r<\eta\;,\\
	\rho_T''(B,\eta,r)\;, \quad \eta<r<2\eta\;,\\
	\frac{6(1-\varepsilon)}{r^4}\;,\quad r>2\eta\;,
\end{cases}
\label{DenBR''}
\ee

\be
 p_\perp'(B,\eta,r)\;=\;
\begin{cases}
	\frac{3}{2}\;,\quad  r<\eta\;,\\
	p_{T\perp}'(B,\eta,r)\;, \quad \eta<r<2\eta\;,\\
	0\;,\quad r>2\eta\;.
\end{cases}\label{PBR'}
\ee

The continuity of $\rho''(B,\eta,r)$ and $p_\perp'(B,\eta,r)$ at $\;r=\eta\;$ under the condition~(\ref{B-cond}) forces us to fix  $\;B=\frac{1}{50}(60-60\varepsilon+57\eta^3)\;$ and further restrains  $\;0<\eta<0.69\;$ and $\;0<\varepsilon<\frac{19}{20}\eta^3\;$. However, continuity at $\;r=2\eta\;$ rather requires  that  $\;B=\frac{1}{76}(120-120\varepsilon+75\eta^3)\;$ and $\;\frac{5}{8}\eta^3<\varepsilon<\frac{1}{200}(48+125\eta^3)\;$.

Our conclusion is that, to satisfy  this amount  of continuity on both ends simultaneously,  necessitates  the inclusion of two more matching conditions, $\;f_T'''(B,\eta,r=\eta)=\frac{6}{4}\;$ and $\;f_T'''(B,\eta,r=2\eta)=0\;$; meaning that our polynomial ansatz~(\ref{Polynom}) would  have to be extended to one of seventh order.

Clearly, enforcing additional constraints on the continuity of even
higher-order  derivatives of the  energy density and the transverse pressure,
one requires a
yet higher-order polynomial ansatz.

\subsection{Graphs}

Here, we present graphs for the metric function, energy density, first derivative of the density and transverse pressure in the core of the frozen star and its vicinity, $\;r< 4\eta\;$, as a function of the radius $r$ in units of $R$. The values for $B$, $\eta$ and $\varepsilon$ are meant as reasonable estimates but otherwise chosen arbitrarily.

\vspace{-.5in}
\begin{figure}[H]
	\hspace{-0.205in}\includegraphics[scale=0.5]{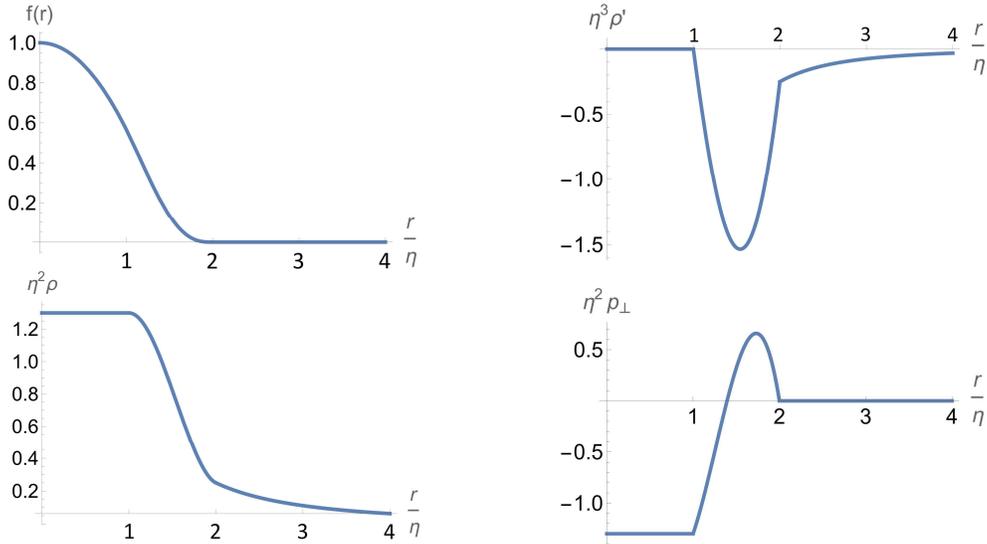}
 	\centering
\vspace{-0.7in}
	\caption{The metric function (upper left), the energy density (lower left), the first derivative of the energy density (upper right) and the transverse pressure (lower right) in the smoothed core of the frozen  for ${B}=1.3$, ${\eta}=0.0001$ and $\varepsilon=0.0001$.}
\end{figure}

\end{appendices}

\end{document}